\documentclass[aps,floatfix,twocolumn,a4paper,noshowpacs, nofootinbib,superscriptaddress,10pt]{revtex4}
\usepackage{graphicx,float}\usepackage{graphicx,float}
\usepackage[all]{xy}
\usepackage{amsmath,upgreek}
\usepackage{amssymb}
\usepackage{color}
\usepackage{epsfig,bm}		
\usepackage{graphicx,epstopdf}
\usepackage{subfigure}
\usepackage{pdfpages,slashed}
\usepackage[colorlinks,hyperindex]{hyperref}

\definecolor{green1}{RGB}{0,128,0}
\hypersetup{backref=true,pagebackref=true}
\hypersetup{%
  colorlinks = true,
  linkcolor  = blue,
  citecolor = cyan,
}
\usepackage{bookmark,textgreek}
\usepackage{hyperref,color,xcolor}
\hypersetup{hidelinks,hyperindex=true,colorlinks=true,breaklinks=true,urlcolor= blue}
\hypersetup{%
  colorlinks = true,
  linkcolor  = blue
}\usepackage{amssymb}
\newsavebox{\foobox}

\usepackage{graphicx,float,tikz}
\usepackage[all]{xy}
\newcommand\ringring[1]{%
  {
   \mathop{\kern0pt #1}\limits^{
     \vbox to-1.85ex{
       \kern-2ex 
       \hbox to 0pt{\hss\normalfont\kern.1em \r{}\kern-.45em \r{}\hss}%
       \vss 
     }
   }
  }
}

\newcommand{\bpartial}{\mathop{\partial\kern -4pt\raisebox{.8pt}{$|$}}}

\newcommand{\bes}{\begin{subequations}}
\newcommand{\ees}{\end{subequations}}
\def\beq{\begin{eqnarray}}

\def\eeq{\end{eqnarray}}
\def\be{\begin{equation}}
\def\ee{\end{equation}}

\begin{document}

\title{Nuclear configurational entropy and high energy hadron-hadron scattering reactions}
\author{G. Karapetyan}
\email{gayane.karapetyan@ufabc.edu.br}
\affiliation{Federal University of ABC, Center of Natural Sciences, Santo Andr\'e, 09580-210, Brazil}
\affiliation{Federal University of ABC, Center of Mathematics, Santo Andr\'e, 09580-210, Brazil}

\begin{abstract}
In this work, the high energy hadron-hadron scattering is studied in the framework of holographic AdS/QCD models, using the Brower-Polchinski-Strassler-Tan Pomeron exchange kernel and gravitational form factors.
We apply the configurational entropy techniques to estimate the   slope of the total cross section for the total hadron-hadron cross sections at high energies.
A good agreement is derived between our approach and the total cross section for combinations that include the pion-nucleon, nucleon-nucleon, and pion-pion, as well as for any high energy data with inclusion of data from the TOTEM collaboration at the LHC and approximated by the Pomeron exchange.
In the case of pion-nucleon and pion-pion scattering, the agreement for the critical points to the differential configurational entropy can be reached within 1.1\% even without the involvement of any extra parameters.
\end{abstract}
\maketitle
\section{Introduction}

The QCD theory of strong interactions is a gauge theory, wherein the Color Glass Condensate (CGC) approach is a  strong and convenient tool to investigate the mechanism of  high-energy processes. In this setup, the differential configurational entropy (DCE) techniques predict the crucial points that help to understand the massive experimental data obtained by different collaborative groups \cite{daRocha:2021jzn,daRocha:2021ntm,Karapetyan:2018yhm,daRocha:2021imz, Fernandes-Silva:2018abr,Ferreira:2020iry, Bernardini:2019stn,Bazeia:2021stz,Bazeia:2021jok,Silva:2021lmy,Bazeia:2021pbs,Bazeia:2020jma,
Braga:2016wzx,Braga:2021fey,daRocha:2021xwq,Braga:2021zyi,Braga:2020hhs}.
At high energy regimes, the interesting and important information about various aspects of the reaction, for example, the total hadron-hadron or photon-hadron cross sections, can be derived by applying the nuclear DCE approach to nuclear systems at extreme conditions \cite{Karapetyan:epjp,Karapetyan:plb, Karapetyan:2020epl, Karapetyan:2021epjp}.
During short time of investigation, the comprehensive studies have been accumulated, in which such approach has been successfully used to describe the mechanism of the nuclear reactions at high energies
\cite{Ferreira:2019inu,Bernardini:2018uuy,Braga:2018fyc,Bernardini:2016hvx,Barbosa-Cendejas:2018mng,Ferreira-Martins:2021cga}, glueballs \cite{Bernardini:2016qit}, charmonium and bottomonium \cite{Braga:2017fsb}, the quark-gluon plasma \cite{daSilva:2017jay}, baryons \cite{Colangelo:2018mrt}, and also using the AdS/QCD model approach \cite{Ma:2018wtw}.
The main details and theory of the CE concept itself based on Shannon's information entropy can be found in Refs.  \cite{Gleiser:2012tu, Gleiser:2011di, Gleiser:2013mga, Gleiser:2018kbq,Sowinski:2015cfa,Gleiser:2014ipa,Gleiser:2015rwa}.
We can also mention other works, where the critical points of the CE have been used to explain the phenomena of AdS black holes and their quantum portrait as Bose--Einstein graviton condensates, also emulating magnetic structures \cite{Casadio:2016aum,Fernandes-Silva:2019fez,Braga:2020myi,Alves:2017ljt,Alves:2014ksa,Bazeia:2018uyg}.
The measure of the minimum values of CE, or the critical points, carries all information content about the stability of the complex nuclear systems. 
In the framework of the CGC, the nuclear DCE critical points have been deeply studied by computing
the inelastic hadron cross sections \cite{Karapetyan:2018yhm,Karapetyan:2019epl,Ma:2018wtw,Karapetyan:plb,Karapetyan:2017edu,Karapetyan:2016fai}. 
It is possible to compute the nuclear DCE when the cross section plays the role of a localized function that is related to the probability of the reaction product for any nuclear spatial configuration.
Hence, one can estimate the critical points of the DCE via the Fourier transform of the cross section and subsequent calculation of the related modal fraction 
\cite{Karapetyan:2018yhm,Karapetyan:plb,Karapetyan:2017edu,Karapetyan:2016fai}.

Recently, the inner composition of hadrons, as well as the quark-gluon pairing, caused a lot of interest because of the numerous experimentally investigated scattering processes which have been done at high energy regime as, for instance, the data taken at LHC. Within the framework of QCD, using the factorization theorem, one can decompose the cross section at high energies into the hard and the soft components.
While the perturbation method of QCD allows estimating  the hard part of the cross section, it is not so straightforward to calculate the soft part due to its not perturbative origin. So, it is just possible to try to use some phenomenological approach, as the parton distribution functions (PDFs) parametrization, which is represented by the Bjorken scaling variable $x$ and the energy scale $Q^2$.
At the high energy regime of hadrons, the nonperturbative approach is one of the most convenient methods to calculate the process cross section within QCD.
In such a context, the holographic QCD and the AdS/CFT correspondence have been used to exam the high-energy scattering reactions. One should suggest the Pomeron exchange (or a multi-gluon exchange) at the small $x$ range to be able to characterize the whole dynamics of the partons.
In Ref. \cite{Brower:2006ea} the deep analysis based on the
gauge/string duality, has been done by Brower, Polchinski, Strassler, and Tan (BPST). This method assumes the contribution of the Pomeron exchange to the cross sections, which can be caused by a kernel in AdS space.

Hence, it is possible to use the BPST kernel approach to calculate the total hadronic cross sections for various scattering processes at a high energy regime. It is worth remembering that the masses of the participants in the hadron-hadron scattering process are the only variables in the AdS/CFT nonperturbative scale.
In the AdS space within the BPST approach, the estimation of the total hadronic cross sections implies the BPST Pomeron exchange kernel and two density distributions.
In the AdS/QCD approach the density distributions are characterized by the gravitational form factors \cite{Watanabe}.
For a non-excited state of hadron, one can use the hard-wall model to describe its behavior. Such a model is valid for AdS geometry in the QCD scale, where there is a cut-off in the infrared region (IR).
In the case of fixed model parameters based on the given experimental data, the cross section for hadron scattering reaction can be easily calculated with the assumption of hadron normalizable mode and its normalized density distribution. Hence, we do not need to use any other parameters and get a universal description of any high-energy process within the Pomeron exchange approximation.

We use the DCE paradigm to investigate the pion-nucleon, nucleon-nucleon, and pion-pion cross section using the Pomeron exchange kernel and gravitational form factors.
Section II is devoted to the Pomeron exchange model in the framework of holographic QCD for high-energy hadronic interactions.
In the next section, we present the main results of the analysis using DCE approach and the concept of the critical points based on the experimentally determined hadron-hadron cross sections.
The summary and outlook are given in Sec. IV.

\section{The total hadron-hadron cross sections within the holographic QCD}

Let us briefly represent the model description in the framework of the PST Pomeron exchange kernel, which allows computing the hadron scattering total cross section ($\chi$).
Within this model the scattering amplitude in eikonal formulation is given by the following expression:
\begin{widetext}
\begin{eqnarray}
{\cal A} (s, t) = 2 i s \int d^2 b e^{i \bm{k_\perp } \cdot \bm{b}} \int dzdz' P_{13}(z) P_{24}(z') \left[ 1-e^{i \chi (s, \bm{b}, z, z')} \right].
\label{eq:amplitude}
\end{eqnarray}
\end{widetext}
In Eq. \eqref{eq:amplitude} the Mandelstam variables are denoted by $s$ and $t$, whereas the two-dimensional impact parameter is denoted by $\bm{b}$. Finally, $z$ and $z'$ are  coordinates in AdS for the incident and target particles, respectively, 
and $P_{13}(z)$ and $P_{24}(z')$ are the density distributions of the two hadrons in the AdS space, which are normalized according to the following conditions:
\begin{equation}
\int dz P_{13}(z) = \int dz' P_{24}(z') = 1,
\label{eq:normalization_condition}
\end{equation}
Taking into account the optical theorem, in the single-Pomeron exchange model the total cross section can be expressed as \cite{Watanabe}
\begin{equation}
\sigma(s) = 2 \int d^2b \int dzdz' P_{13} (z) P_{24} (z') \Im \chi (s,\bm{b},z,z'). \label{eq:tcs_original}
\end{equation}
The final expression for the total hadron-hadron cross section can be written as \cite{Watanabe}
\begin{align}
&\sigma(s) = \frac{g_0^2 \rho^{3/2}}{8 \sqrt{\pi}} \int dzdz' P_{13} (z) P_{24}(z') (zz') \Im [\chi_{c}(s,z,z')], \label{eq:tcs_with_CK} \\
&\Im [\chi_c(s,z,z') ] \equiv\frac{1}{\sqrt{\tau}} \exp\left[{(1-\rho)\tau-\frac1{\rho \tau}\log ^2\left(\frac{z}{z'}\right)}\right], \label{eq:CK}
\end{align}
where ${Im} [\chi_{c}(s,z,z')]$ i s the imaginary part of the scattering amplitude and $\tau = \log (\rho z z' s/2)$ with $g_0^2$ and $\rho$  are the overall factor and the slope of the total cross section, respectively.
It should be mentioned that in the nucleon-nucleon scattering the main value that describes the process is the nucleon mass ($m_N \sim 1$ GeV). This fact supports the idea the leading role of the
the low energy dynamics with a strong coupling in QCD.
It means that one can use the modified BPST kernel with the same functional form as the conformal kernel in the following form,
\begin{align}
&\Im [\chi_{mod} (s, z, z')]=
\Im [\chi_c (s, z, z') ]\nonumber\\&\qquad\qquad\qquad\qquad+ \mathcal{F} (s, z, z') \Im [\chi_c (s, z, z_0 z_0' / z') ],
\label{eq:MK}
\end{align}
where
\begin{align}
\mathcal{F} (s, z, z') &= 1 - 2 \sqrt{\rho \pi \tau} e^{\eta^2} \mbox{erfc}( \eta ), \\
\eta &=  \frac{-\log \left(\frac{z z'}{z_0 z_0'} \right) + \rho \tau}{\sqrt{\rho \tau}},
\end{align}
with the coordinates cut-off parameters, $z_0$ and $z'_0$, in QCD scale, fixed by hadron masses.

The density distributions, $P_{13}(z)$ and $P_{24}(z')$, of the interacting hadrons in Eq. \eqref{eq:amplitude} are described by the gravitational form factors. Such form factors can be received from the bottom-up AdS/QCD models using the hadron-Pomeron-hadron three-point functions.
As the nucleon can be expressed as a solution to the five-dimensional Dirac equation \cite{Watanabe}, the density distribution of the nucleon is expressed with the left-handed and right-handed components of the Dirac field ($\psi_L$ and $\psi_R$) via the Bessel function in the following form:
\begin{align}
&P_N (z) = \frac{1}{2z^{3}} \left[ \psi_L^2 (z) + \psi_R^2 (z) \right], \\
&\psi_L (z) = \frac{\sqrt{2} z^2 J_2 (m_N z)}{z_0^N J_2 (m_N z_0^N)}, \
\psi_R (z) = \frac{\sqrt{2} z^2 J_1 (m_N z)}{z_0^N J_2 (m_N z_0^N)},
\end{align}
The fixed cut-off parameter ($z_0^N=4.081 keV$) obeys the conditions of the normalization $J_1 (m_N z_0^N) = 0$ with $m_p$ and $m_n$ being the proton and neutron physical masses, respectively.
The pion wave function $\Psi$ is extracted from the bottom-up AdS/QCD model motion equation of mesons \cite{Watanabe}. Hence, we can present the density distribution of the pion in the form of the Bessel function as the following:
\begin{align}
&P_\pi (z) = \frac{ \left[ \Psi'(z) \right] ^2 }{4 \pi^2 f_\pi ^2 z}  + \frac{\sigma^2 z^6 \Psi (z)^2 }{ f_\pi ^2 z^3 },
\end{align}
where the prime denotes the derivative with respective to $z$ and the pion wave function reads 
\begin{widetext}
\begin{align}
&\Psi \left( z \right) = z\Gamma \left(\frac{2}{3} \right) \left( {\frac{\alpha }{2}} \right)^{1/3} \Biggl[ I_{ - 1/3} \left( {\alpha z^3 } \right) - I_{1/3} \left( {\alpha z^3 } \right)\frac{{I_{2/3} \left( {\alpha (z_0^\pi)^3 } \right)}}{{I_{ - 2/3} \left( {\alpha (z_0^\pi)^3 } \right)}} \Biggr]
\end{align}
\end{widetext}
 with $z_0^\pi = 3.105 {\rm keV}$ and $\alpha = \frac{2 \pi}3 \sigma$, for $\sigma = (332$ MeV$)^3$, and $f_\pi$ is the pion decay constant.
The cutoff parameter, $z_0^\pi$, encodes the mass of $\rho$ meson ($m_\rho$) by the zeroes of the zeroth order Bessel function as $J_0 (m_\rho z_0^\pi) = 0$.

\section{Configurational entropy via the BPST Pomeron exchange kernel}

To determine the density distribution, one does not need any additional adjustable parameter. Within the BPST Pomeron exchange kernel approach, the only two parameters that should be extracted from the experiments are $g_0^2$ and $\rho$ (see Eq. \eqref{eq:CK}).
During our analysis, we use the experimental data of TOTEM collaboration at LHC \cite{Watanabe}, as well as the recent hadron-hadron collision data which are given in the Particle Data Group within the energy range $10^2 < \sqrt{s} < 10^5$ GeV.
As the check point values for the two adjustable parameters have been used the data from Ref. \cite{Watanabe}, $g_0^2 = 6.27 \times 10^2$ and $\rho = 0.824$.

We use the Pomeron exchange kernel to calculate the total hadron-hadron scattering cross sections at a high energy range.
As the using adjustable parameters of the model do not depend on the intrinsic properties of hadrons, one can assume the Pomeron exchange approach is a universal tool to describe the process at high energies.
This fact is supporting also by the chosen normalizable modes of both the nucleon and the pion.
In this way, one can easily compute the cross section for the incident pion, if the parameters for the nucleon-nucleon interaction are already fixed, and do not include other adjustable parameters.
Besides, the scale of the scattering processes at high energies is characterized by the constant masses of the participated hadrons. Therefore, in such a case, we also have the same conditions for all three processes under study.
which leads to the constant value for the total cross section ratios among them.
In this paper, we use the normalized structure factor for the collective coordinates, which have been obtained by the Fourier function decomposition into the number of weighted components for the total hadron-hadron cross sections \cite{Bernardini:2016hvx}.
The cross section for any collision, as the probability of the nuclear reaction, is governed by the modal fraction of the CE, and, consequently, contains all the needed information.
On the other hand, the limit number of the critical points that can be determined using the Shannon entropy and the configurational entropy concepts, totally characterize an isolated localized nuclear system \cite{Gleiser:2012tu,Gleiser:2013mga}.

If one considers as a localized function the reaction cross section, it is possible to calculate the latter by via the Fourier transformation energy--weighted correlation function, given by the cross section, 
\begin{equation}
\label{34}
\sigma_{}({k,\rho})=\frac{1}{2\pi}\! \int_{\mathbb{R}}\sigma_{} (z,\rho)\, e^{ikz} d z.
\end{equation}
Then, the modal fraction of CE can be determined as:
\begin{equation}\label{modall}
f_{\sigma_{}({k,\rho})}=\frac{\vert \sigma_{}({k,\rho}) \vert^2}{\int_{\mathbb{R}}\vert \sigma_{}({k,\rho})\vert ^2 dk}.
\end{equation}
The set of the critical points is determined through the suitable expression for the DCE \cite{Gleiser:2012tu}:
\begin{equation}
\label{333}
{\rm CE}(a,b) = - \int_{\mathbb{R}} f_{\sigma_{}({k,\rho})} \log f_{\sigma_{}({k,\rho})}d k.
\end{equation}
And finally, one obtains the CE critical points just by applying Eqs. (\ref{34} - \ref{333}) employing the total hadronic cross section.

\begin{figure}[h]
	\centering
	\includegraphics[width=6.8cm]{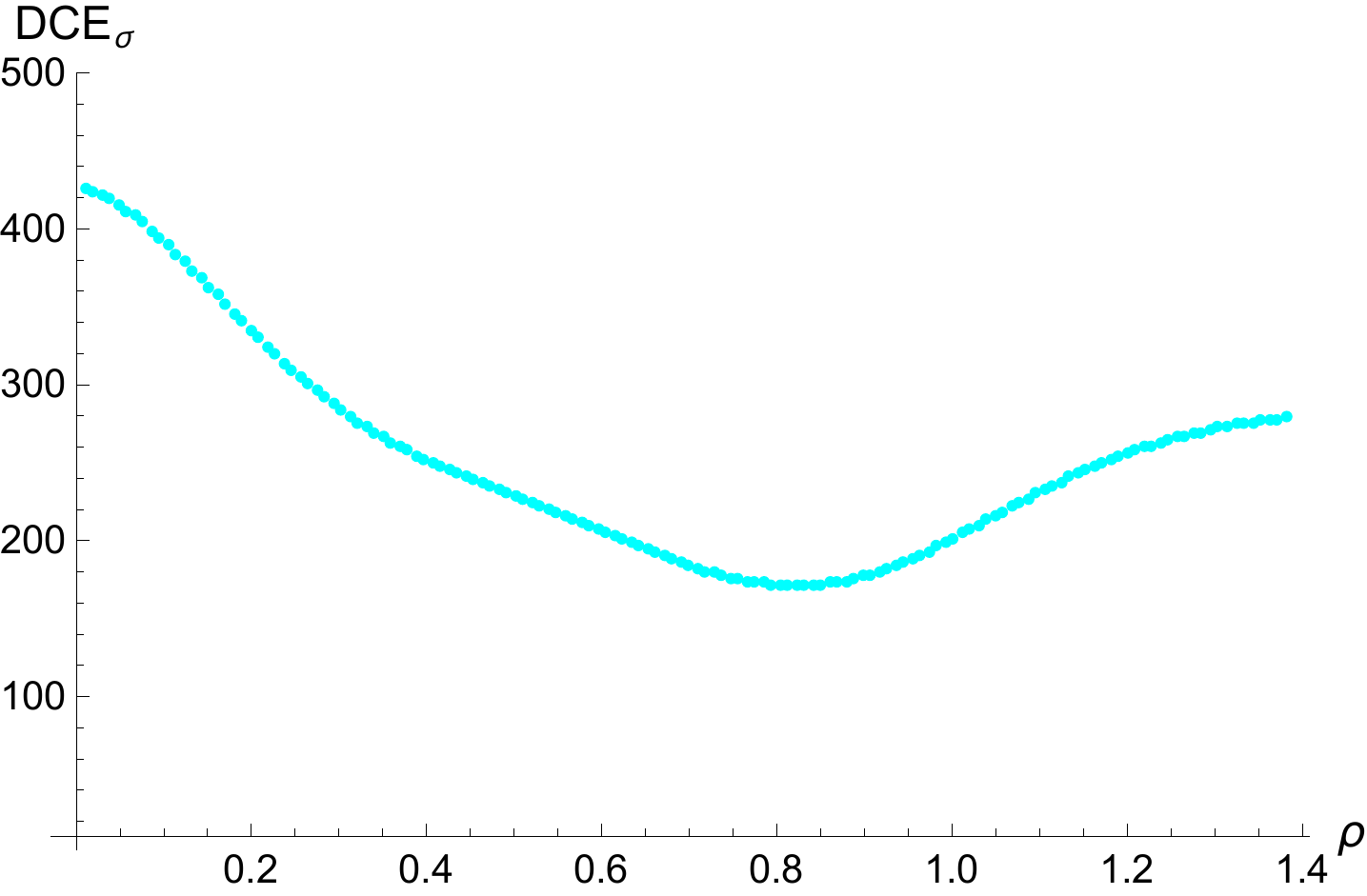}
	\caption{DCE $\times$ $\rho$. Minimum: $\rho = 0.814$, DCE = 164.72 nat (Natural unit of information.)}
	\label{fi1}
\end{figure}

In Fig. \ref{fi1} one can see the results of the DCE versus $\rho$ calculations for the total hadron-hadron scattering cross sections at a high energy regime.
Fig. \ref{fi1} shows the predominance of the
quantum states with the slope of the total cross section $\rho = 0.814$, corresponding to the nuclear DCE = 164.72 nat, with the corresponding global minimum of the nuclear configurational entropy at such point.
One can see that the differential configurational entropy curve tends to grow on both sides from the observed global minimum, although with different signs when one looks at the second derivative of the nuclear DCE with respect to $\rho$, in Fig. \ref{fi1}.
The result of our calculation for hadron-hadron interactions at high energies fits well with the value of parameter $\rho=0.824$ obtained in Ref. \cite{Watanabe}, within 1.1\%.
The critical point of the nuclear DCE indicates the natural choice of the total hadron cross section for the parameter, which controls the energy dependence of the cross sections and confirms the earlier received experimental data with high enough precision. At  $\rho = 0.814$, the nuclear system 
stays at its most stable configuration. 
Such minimum has been computed through Shannon's information entropy approximation \cite{Gleiser:2011di} and contains all information about the mechanism of the reactions.
The fitting procedure of adjustable parameter for the experimentally determined total hadron-hadron cross sections clearly shows the conditions for the configurational stability of the excited nuclear system at high energies.
The stability of any localized nuclear configuration is characterized by the set of limited number degrees of freedom, which determine the most dominant state(s) after reaching the thermodynamic equilibrium of the system. And our results once again confirm this fact.
Of course, each case of the reaction requires special and deeper investigation using the comprehensive results obtained via the recent studies in this area \cite{Bazeia:2013usa,Correa:2015lla,Correa:2015vka}

\section{Conclusions}

The total cross sections for three hadron-hadron scattering reactions at a high energy regime have been studied within the Pomeron exchange kernel model of holographic QCD \cite{MarinhoRodrigues:2020yzh,MarinhoRodrigues:2020ssq}.
This model allowed us to calculate the total hadron cross sections using the expression for the density distributions for the nucleon and for pion, which are expressed by the gravitational form factors in the framework of bottom-up AdS/QCD models.
The model conditions imply complete independence from the observed $s$ quantity because of the energy dependence of the cross section controlled by the BPST kernel.
In the considering hadron energy range ($10^2 < \sqrt{s} < 10^5$ GeV) the total scattering cross sections are in satisfactory agreement with the data from Ref. \cite{Watanabe}.
We can conclude based on the results of this paper that the BPST Pomeron exchange kernel is a convenient tool in the investigation of high energy hadron interactions with nonperturbative gluonic dynamics.
We applied the nuclear DCE approach and estimated its  global minimum, which characterizes the critical and stable point of the hadron-hadron reactions.
The calculation of the total scattering cross section within the BPST Pomeron exchange kernel determined the natural choice for the excited nuclear configuration.
We found the satisfactory agreement of the calculated parameters with those predicted by \cite{Watanabe}, within 1.1\%.
The stability of the nuclear system can be completely determined by the critical point of the nuclear DCE. Hence, all the available information can be reached through the systematic study of the DCE approach.

\paragraph*{Acknowledgments:} GK thanks to The S\~ao Paulo Research Foundation -- FAPESP (grant No. 2018/19943-6).%

\end{document}